\documentclass[conference,a4paper]{IEEEtran}
\IEEEoverridecommandlockouts
\usepackage[english]{babel}
\addto\captionsenglish{}
\usepackage[utf8]{inputenc}
\usepackage{cite}
\usepackage{amsmath,amssymb,amsfonts}
\usepackage{graphicx}
\usepackage{textcomp}
\usepackage{xcolor}
\def\BibTeX{{\rm B\kern-.05em{\sc i\kern-.025em b}\kern-.08em
    T\kern-.1667em\lower.7ex\hbox{E}\kern-.125emX}}
\usepackage{subfig}
\usepackage{siunitx}
\usepackage{xspace}
\usepackage{etoolbox}
\usepackage{booktabs}
\usepackage{hyperref}
\usepackage{bbm}
\usepackage{eso-pic}

\AddToShipoutPicture*{\footnotesize\sffamily\raisebox{1cm}{\hspace{1.65cm}\fbox{\parbox{\textwidth}{\copyright~2019 IEEE. Personal use of this material is permitted. Permission from IEEE must be obtained for all other uses, in any current or future media, including reprinting/republishing this material for advertising or promotional purposes, creating new collective works, for resale or redistribution to servers or lists, or reuse of any copyrighted component of this work in other works.}}}}

\newcommand*{\figref}[1]{Fig.~\ref{#1}}
\newcommand*{\secref}[1]{Section~\ref{#1}}
\newcommand*{\tabref}[1]{Table~\ref{#1}}
\newcommand*{\agilent}{Agilent\xspace}
\newcommand*{\keysight}{Keysight\xspace}
\DeclareSIUnit\dbm{\decibel{}m}

\begin{document}

\title{3D Field Simulation Model for Bond Wire On-Chip Inductors Validated by Measurements
}

\author{\IEEEauthorblockN{Katrin Hirmer and Klaus Hofmann}
\IEEEauthorblockA{\textit{Integrated Electronic Systems Lab} \\
\textit{Technische Universität Darmstadt, Germany}\\
Merckstr. 25, 64283 Darmstadt\\
\{Katrin.Hirmer,Klaus.Hofmann\}@ies.tu-darmstadt.de}
\and
\IEEEauthorblockN{Thorben Casper and Sebastian Schöps}
\IEEEauthorblockA{\textit{Computational Electromagnetics} \\
\textit{Technische Universität Darmstadt, Germany}\\
Dolivostr. 15, 64293 Darmstadt\\
\{casper,schoeps\}@temf.tu-darmstadt.de}
}

\maketitle

\begin{abstract}
This paper proposes 3D field simulation models for different designs of integrated bond wire on-chip inductors. To validate the simulation models, prototypes for three designs with air and ferrite cores are manufactured and measured. For air core inductors, high agreement between simulation and measurement is obtained. For ferrite core inductors, accurate models require an exact characterization of the ferrite material. These models enable the prediction of magnetic field influences on underlying integrated circuits.
\end{abstract}


\section{Introduction}

Inductors are critical components in most power electronic and radio-frequency (RF) circuits. However, their integration with the goal of high inductances on small areas remains an ongoing research topic~\cite{dinulovic2018high}. A common approach for RF integrated circuits (ICs) is to realize inductances by 2D integrated planar spiral inductors~\cite{mohan1999}. These inductors are easy to implement and compatible for every process.
However, this approach requires a high area consumption~\cite{yook2005high}.
Moreover, the magnetic flux is perpendicular to its substrate causing possible interferences with the circuit within the IC.

For power applications, high inductances and high quality-factors (Q-factors) are required.
Thus, the trend to miniaturization leads to an increasing research on power supply on chip (PwrSoC) with focus on inductors and transformers~\cite{mathuna2012,dinulovic2018high,dinulovic2016chip}.
To obtain the required high inductances for power applications, discrete semiconductors are more promising than 2D spiral integrated inductors~\cite{mathuna2012}.
In particular, the fabrication of on-chip inductors and transformers using bond wires around a magnetic core is a promising approach~\cite{sullivan2013,camarda2018design,macrelli2015modeling}.
Since the bonding process is a common subsequent process step after IC fabrication, this approach results in low additional costs and a low production time~\cite{Shen2007}.
In addition, these components are less area consuming yielding a high power density and efficiency.

\begin{figure}
    \centering
    \subfloat[][]{\includegraphics[width=0.4\linewidth]{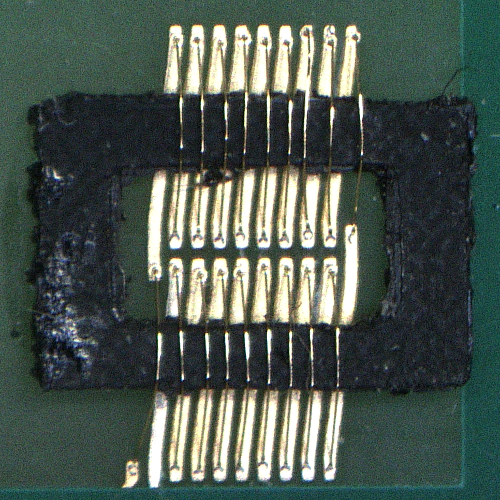}}
    \qquad
    \subfloat[][]{\includegraphics[width=0.4\linewidth]{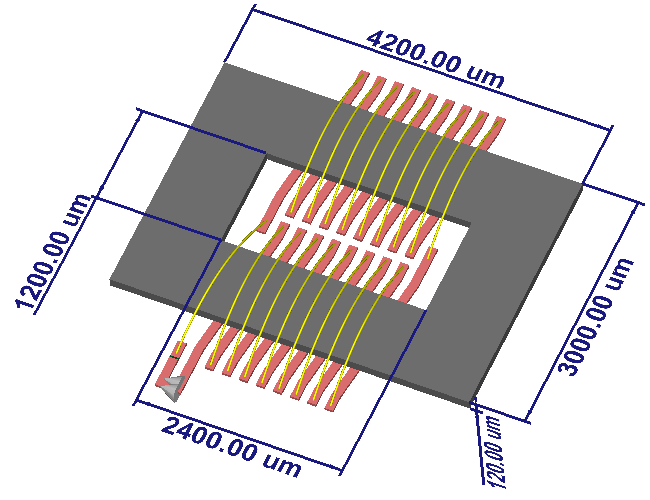}}
    \caption{(a) Prototype and (b) simulation view of a closed core magnetic inductor (model C).}
    \label{fig:modelC}
\end{figure}

To reduce the number of prototypes, IC designers rely on simulation tools that require reliable inductor and transformer models.
This work proposes bond wire on-chip inductors to be designed with the help of 3D field simulation. To test different geometric and material parameters without having to fabricate additional prototypes, simulation models were set up and validated using measurements of corresponding prototypes, such as the one in \figref{fig:modelC}.

The paper is structured as follows.
First, \secref{sec:simulation} introduces the simulation model to yield the resistance and inductance of an inductor.
Then, \secref{sec:measurement} presents the prototypes and the measurement setup.
\secref{sec:results} discusses the results focusing on the comparison between measurement and simulation and \secref{sec:conclusions} concludes the paper.

 \section{Simulation Model}
\label{sec:simulation}

To avoid the costly fabrication of prototypes, simulation models were set up to predict the devices' behavior.
In this paper, the major quantities of interest are the resistance and the inductance of the on-chip inductor.
This allows to consider appropriate stationary simplifications of Maxwell\xspace's equations~\cite{jackson1998}.
First, to compute the DC resistance, one needs to solve the electrokinetic problem
\begin{equation}
    \ensuremath{\nabla\cdot}\xspace(\sigma(\ensuremath{\mathbf{x}}\xspace)\ensuremath{\nabla}\xspace\varphi(\ensuremath{\mathbf{x}}\xspace))=0,
    \label{eq:Jcont}
\end{equation}
with magnetic boundary conditions and, e.g., $\varphi=0$ and $\varphi=V$ at the contacts.
Here, $\varphi$ is the electric scalar potential and $\sigma$ is the electric conductivity.
Then, the current $I=-\int_{A}\sigma\ensuremath{\nabla}\xspace\varphi\ \mathrm{d}\ensuremath{\mathbf{x}}\xspace$ can be computed, where $A$ is the cross-sectional area of a bond wire.
Therefore, the DC resistance yields $R=V/I$.
Secondly, the inductance can be calculated from the magnetostatic field that is obtained from
\begin{equation}
    \ensuremath{\nabla\times}\xspace\left(\nu(\ensuremath{\mathbf{x}}\xspace)\ensuremath{\nabla\times}\xspace\ensuremath{\vec{A}}\xspace(\ensuremath{\mathbf{x}}\xspace)\right)=\chi(\ensuremath{\mathbf{x}}\xspace)I,
    \label{eq:curlcurl}
\end{equation}
where $\nu$ is the reluctivity, \ensuremath{\vec{A}}\xspace is the magnetic vector potential, and $\chi$ is a winding function that distributes the current $I$ accordingly.
From \ensuremath{\vec{A}}\xspace, the magnetic field \ensuremath{\vec{H}}\xspace and the magnetic flux density \ensuremath{\vec{B}}\xspace are obtained by $\ensuremath{\vec{B}}\xspace=\nu^{-1}\ensuremath{\vec{H}}\xspace=\ensuremath{\nabla\times}\xspace\ensuremath{\vec{A}}\xspace$.
Then, the magnetic energy $W$ is calculated by
\begin{equation*}
    W=\frac{1}{2}\int_{\Omega}\vec{B}(\ensuremath{\mathbf{x}}\xspace)\cdot\ensuremath{\vec{H}}\xspace(\ensuremath{\mathbf{x}}\xspace)\ \mathrm{d}{}\ensuremath{\mathbf{x}}\xspace,
\end{equation*} 
where $\Omega$ is the volume in which the energy shall be evaluated, and the inductance reads $L=2W/I^{2}$.
Therefore, computing the resistance and the inductance of an on-chip inductor boils down to solving \eqref{eq:Jcont} and \eqref{eq:curlcurl}.
For this purpose, \textsc{CST EM STUDIO}\xspace is used.
 \section{Prototypes and Measurement Setup}
\label{sec:measurement}

\begin{figure}
    \centering
    \subfloat[\label{fig:modelA}]{\includegraphics[width=0.4\linewidth]{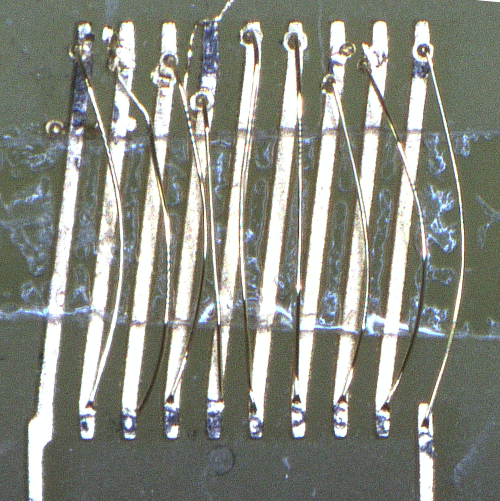}\qquad
                                  \includegraphics[width=0.4\linewidth]{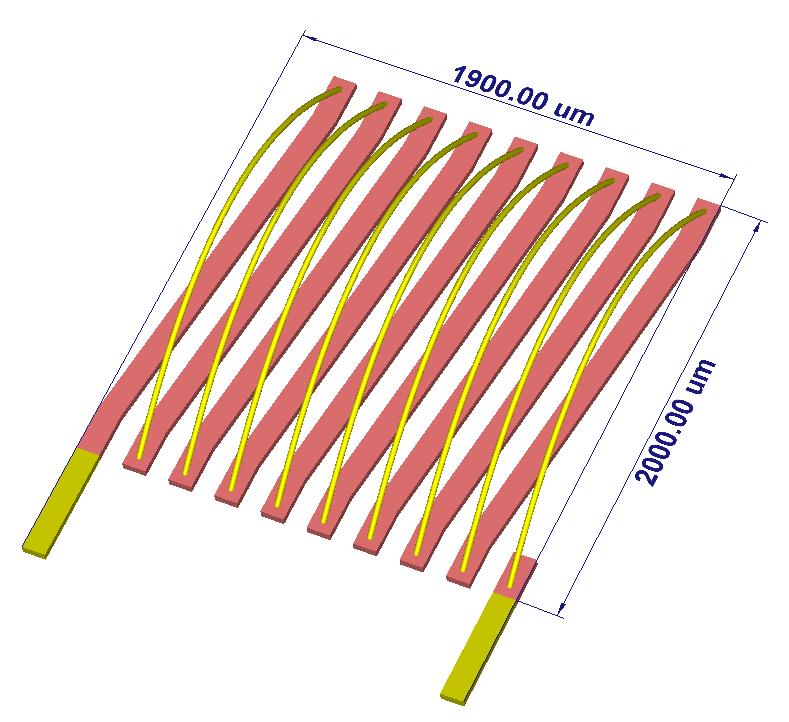}}\\%
    \centering
    \subfloat[\label{fig:modelB}]{\includegraphics[width=0.4\linewidth]{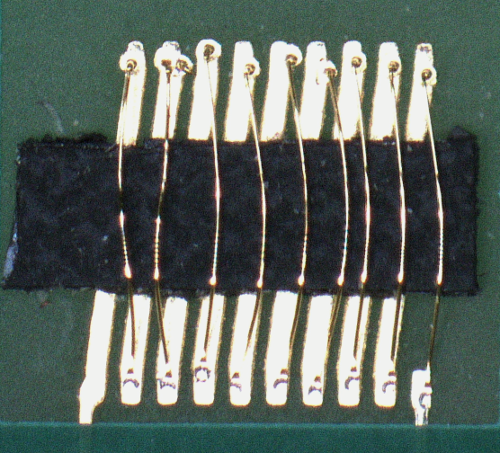}\qquad
                                  \includegraphics[width=0.4\linewidth]{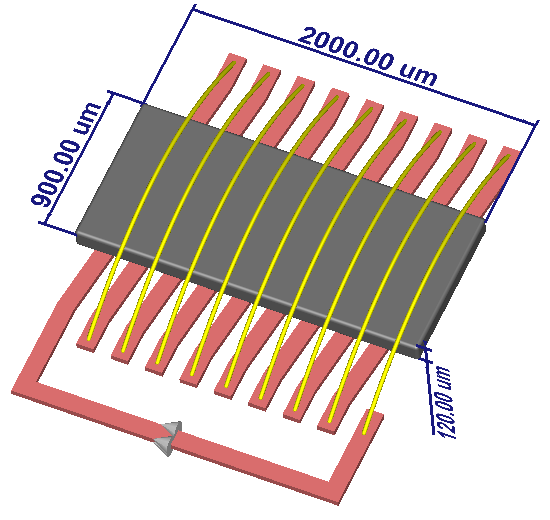}}
    \caption{Prototype and simulation model of (a) an air core inductor (model~A) and (b) a ferrite core inductor (model~B).}
\end{figure}

Three different bond wire inductor models and one bond wire transformer model were fabricated on a printed circuit board (PCB).
Model~A is an air core solenoid inductor with $N=9$ windings and a total footprint of ${A=\SI{2.0x1.9}{\milli\metre}}$, see \figref{fig:modelA}.
Model~B differs from model~A only in an additional ferromagnetic core material as shown in \figref{fig:modelB}.
Model~C consists of a closed ferromagnetic core with two bond wire inductors in series that are placed on opposite legs of the ferromagnetic core.
Each of the two inductors consists of $N=9$ windings and model~C has a total footprint of ${A=\SI{4.2x3.0}{\milli\metre}}$ as shown in \figref{fig:modelC}. 
\tabref{tab:Fabrication} summarizes the different parameters for the three models.
Finally, the transformer model was realized by omitting the wires that connects the two serial inductors in model C.

\begin{table}
\caption{Fabrication parameters for the different solenoid inductor models. Here, $N$ is the number of windings and $A$ is the size of the footprint.}
        \centering
        \begin{tabular}{cccc}\toprule
        Model & $N$ & $A$ in \si{\milli\metre\squared} & Core\\ \bottomrule\toprule
        Model A & 9 & 3.8 & Air \\
        Model B & 9 & 3.8 & Bar ferrite \\ 
        Model C & 18 & 16.8 & Closed ferrite \\ \bottomrule
        \end{tabular}
        \label{tab:Fabrication}
\end{table}

The solenoid inductors and the transformer were fabricated using gold wires with a diameter of \SI{25}{\micro \meter} on a \SI{1.55}{\milli \meter} PCB with \SI{35}{\micro\meter} thick and \SI{100}{\micro\meter} wide copper traces and an electroless nickel electroless palladium immersion gold (ENEPIG) surface plating.
For models B and C, the magnetic core consists of a ferrite foil which was trimmed using a laser cutter.
Its initial relative permeability is given by $\mu_{\text{r;i}}=150\pm 20\%$ at \SI{13.56}{\mega\hertz}~\cite{MULL5040}.

\begin{figure}[!ht]
    \centering    
    \includegraphics{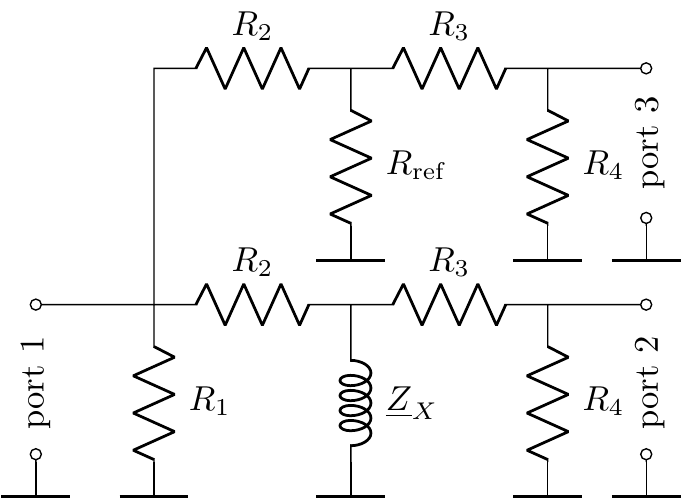}
    \caption{Test setup for differential measuring of the solenoid inductors.}
    \label{fig:micro_setup}
\end{figure}

The impedance of the inductors were measured by an \agilent E5062A network analyzer which is capable of measuring S-parameters up to \SI{3}{\giga\hertz}.
Input and output reflections are reduced by a matching network built around the inductor (device under test) $\underline{Z}_{X}$, see \figref{fig:micro_setup}.
To avoid reflections, resistor values of ${R_{1}=\SI{62}{\ohm},\ R_{2}=R_{3}=\SI{470}{\ohm}}$, and $R_{4}=\SI{58}{\ohm}$ were chosen such that an input and output impedance of \SI{50}{\ohm} was obtained.
To compensate parasitic influences of the PCB and the cables, a differential measurement method was used.
Therefore, a matching circuit with a reference resistor $R_{\text{ref}}=\SI{5}{\ohm}$ is connected in parallel to the test channel.

Relating the measured forward voltage gain $\underline{S}_{21}$ at the test channel to the one at the reference channel denoted by $\underline{S}_{31}$, the impedance $\underline{Z}_{X}$ is determined by
\begin{equation}
    \underline{Z}_{X} = R_{\text{ref}}\frac{\underline{S}_{21}}{\underline{S}_{31}}.
\end{equation}

For frequencies far below the resonance frequency of the coil, the inductance can be calculated by
\begin{equation}
    L_{X}(f)=\frac{\text{Im}(\underline{Z}_{X})}{2 \pi f}.
    \label{eqn:inductance}
\end{equation}
The Q-factor is given by~\cite{Brander2008}
\begin{equation}
    Q(f)=\frac{\text{Im}(\underline{Z}_{X})}{\text{Re}(\underline{Z}_{X})}.
    \label{eqn:q-factor}
\end{equation}
The DC resistances $R_{\text{DC}}$ of the prototypes were measured with a 4-wire sensing \keysight 34465A multimeter.  

 \section{Results and Discussion}
\label{sec:results}

\begin{figure}
    \centering
    \subfloat[\label{fig:micro_im}]{\includegraphics{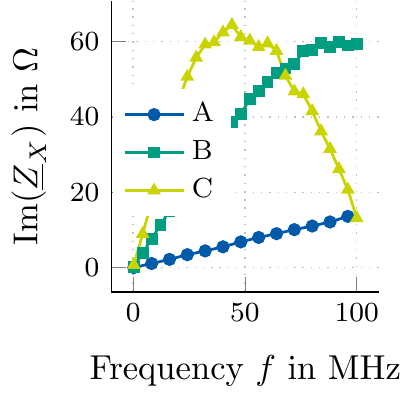}}
    \hspace{-1em}
    \subfloat[\label{fig:micro_re}]{\includegraphics{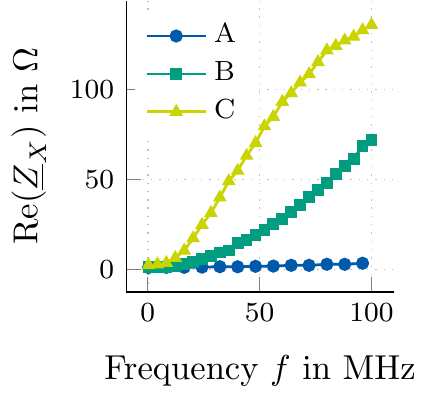}}
    \caption{Measured (a) imaginary and (b) real impedances of models A, B, and C.}
    \label{fig:microImpedance}
\end{figure}

\begin{figure}
    \centering
    \subfloat[\label{fig:micro_l}]{\includegraphics{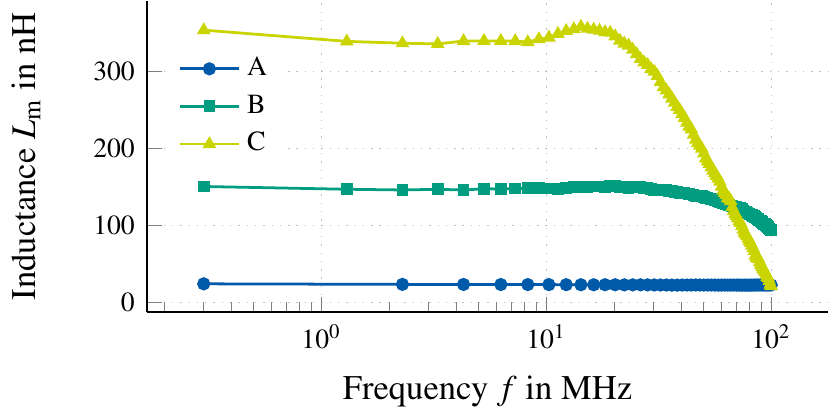}}\\
    \subfloat[\label{fig:micro_q}]{\includegraphics{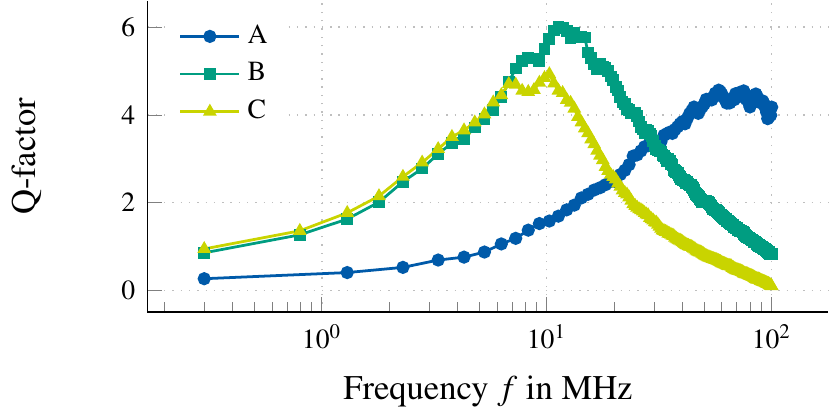}}
    \caption{(a) Inductances $L_{\text{m}}$ and (b) Q-factor extracted from the measurements of models A, B, and C with respect to frequency.}
    \label{fig:micro_l}
\end{figure}

In this section, the results of the measurements and simulations are presented.
First, the complex impedance $\underline{Z}_{X}$ has been measured at an input power of \SI{0}{dBm}.
It is shown for the three different models with respect to frequency in \figref{fig:microImpedance}.
As expected, the frequency dependent losses of the core material~\cite{MULL5040} result in an increase of Re($\underline{Z}_{X}(f)$) at higher frequencies for models B and C.
On the other hand, only a slight increase with frequency is observed for the coreless model A.
The inductances and Q-factors were extracted from S-parameter measurements and are shown with respect to frequency in \figref{fig:micro_l} and \figref{fig:micro_q}, respectively.
The ferrite core of models B and C leads to a significant increase in inductance, see \figref{fig:micro_l}.
However, due to the losses in the core, only a small increase of the Q-factor is observed, see \figref{fig:micro_q}.

\begin{figure}
    \centering
    \includegraphics{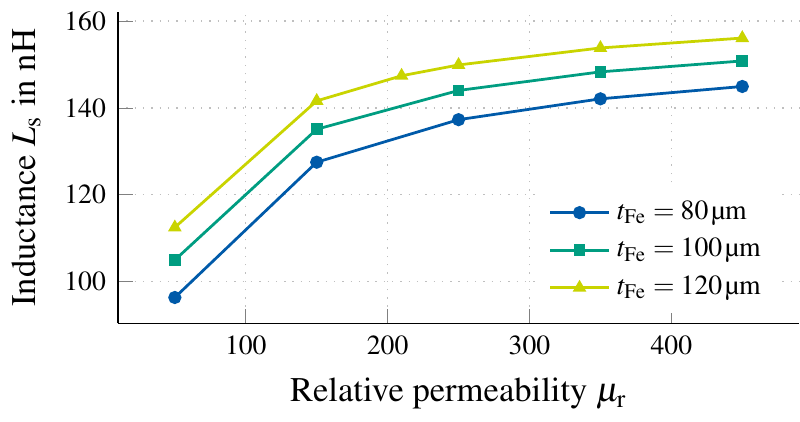}
    \caption{Simulated inductance $L_{\text{s}}$ of model B with respect to the core's thickness $t_{\text{Fe}}$ and its relative permeability $\mu_{\text{r}}$.}
    \label{fig:Sim_B}
\end{figure}

\begin{figure}
    \centering
    \subfloat[][]{\includegraphics[width=0.5\linewidth]{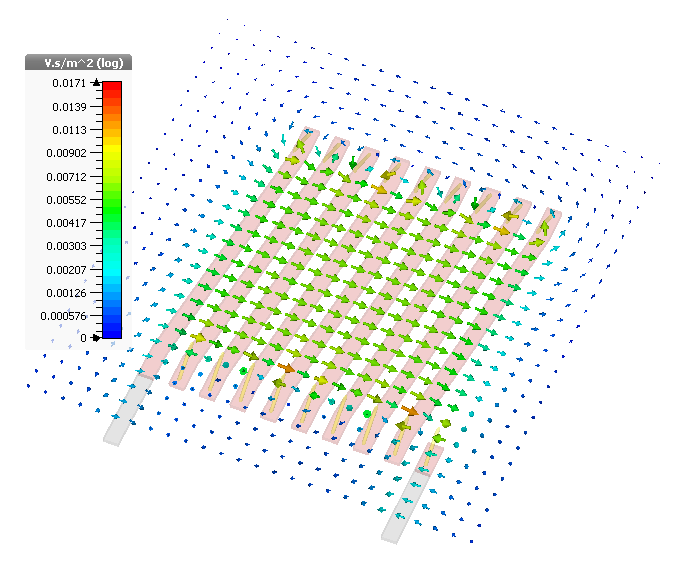}}
    \subfloat[][]{\includegraphics[width=0.5\linewidth]{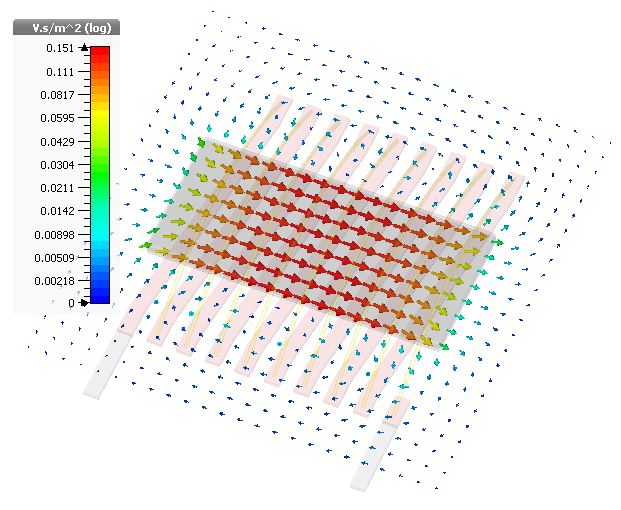}}
    \caption{Magnetic flux density of (a) model A and (b) model~B.}
    \label{fig:fieldplots}
\end{figure}

For the simulations, \textsc{CST EM STUDIO}\xspace was used to solve \eqref{eq:Jcont} and \eqref{eq:curlcurl} and to compute the resistance and the inductance of the three different models.
In the simulation of all considered models, electric boundary conditions were applied at a distance of \SI{6}{mm} from the model itself and a current of \SI{1}{A} was applied as the excitation.
To take into account the uncertainty in the core's material parameters~\cite{MULL5040}, a parameter sweep for model B was carried out to compute the inductance for the nominal and for the corner values of the core's thickness.
In \figref{fig:Sim_B}, the results for model B are shown with respect to the relative permeability.
Fixing the core's thickness to a value of \SI{120}{\micro\metre}, a numerical optimization was applied to adjust the simulation model to the measurement results for model B.
The optimization yielded a permeability of $\mu_{\text{r}}=\num{209.83}$ resulting in a simulated inductance of \SI{147.42}{\nano\henry} with a relative error of \SI{0.11}{\percent} with respect to the measured value.
With these values for the thickness and the permeability, model C and the transformer have been evaluated.
To allow an assessment of the parasitic influences of the inductors, \figref{fig:fieldplots} shows the field plots for the magnetic flux density \ensuremath{\vec{B}}\xspace for models A and~B.

\begin{table*}
    \renewcommand{\arraystretch}{1.3}
    \caption{Measured (index m) and simulated (index s) results for the solenoid inductors. The measured inductance has been extracted at \SI{10}{MHz}.}
    \centering
    \begin{tabular}{ccccccccc}\toprule
        Model & 
        $L_{\text{m}}$ in \si{\nano \henry} & 
        $L_{\text{s}}$ in \si{\nano \henry} &
        $\epsilon(L)$ in \si{\percent} & 
        $R_{\text{m}}$ in \si{\ohm} &
        $R_{\text{s}}$ in \si{\ohm} &
        $\epsilon(R)$ in \si{\percent} &  
        $Q_{\text{Max}}$ &
        $f_{Q_{\text{Max}}}$ in \si{\mega \hertz}\\
        \bottomrule\toprule
        Model A &23.74& 23.77 &0.13& 1.08 &0.89&17.59&4.63&62.11\\
        Model B &147.26& 147.42 &0.11&  1.12 &0.89&20.54&6.23&11.27\\
        Model C &347.27& 1155.72 &232.80&  2.30 &1.83&20.44&5.18&10.27\\\bottomrule
    \end{tabular}
    \label{tab:meas_micro}
\end{table*}

The results from measurement and simulation are summarized in \tabref{tab:meas_micro}.
The relative errors between measured and simulated resistances are all below \SI{25}{\percent}.
The remaining errors are assumed to stem mainly from the soldering joints that are not modeled in the simulation.
While the relative errors of the inductance of models A and B are below \SI{1}{\percent}, the simulated value for model C is much higher than the measured value.
One possible reason for this deviation stems from non-idealities such as edges resulting from the laser cutting process.
However, the major modeling error is assumed to result from a weak coupling within the sintered magnetic core due to a uniaxial anisotropy of the ferromagnetic foil, as also observed for sputtered ferrite cores in~\cite{wright2010}.

\begin{figure}
    \centering
    \includegraphics{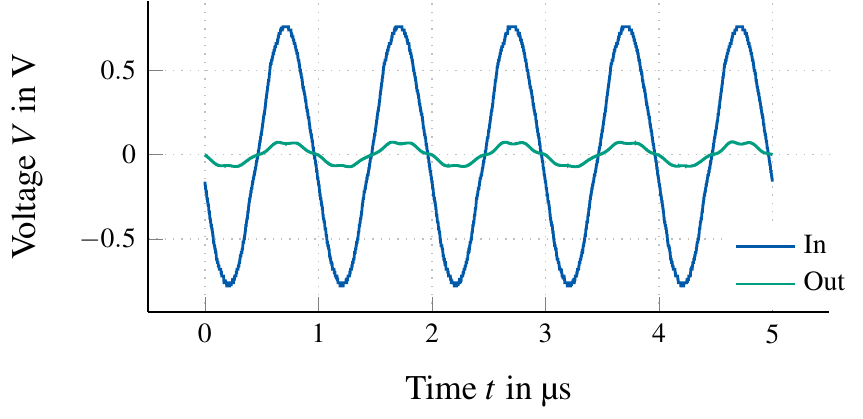}
    \vspace{-2ex}
    \caption{Measured input and output voltage of a 1:1 transformer with respect to time at an input power of \SI{-20}{\dbm}.}
    \label{fig:trafo_V}
\end{figure}

Despite the assumed weak coupling, a 1:1 transformer was successfully built.
The measured input and output voltages for this transformer are shown in \figref{fig:trafo_V} giving a coupling factor of approximately \num{0.10} at \SI{1}{MHz}, whereas the simulated value for ideal isotropy is about \num{0.34}.

\label{sec:conclusions}

Prototypes for on-chip inductors and transformers have been fabricated on printed circuit boards for cost-effective and reliable simulation models which enable system simulation for IC designers.
Measurements of the air core inductor matched well with the simulation results.
For inductances with closed magnetic core, the anisotropy of the material is assumed to be of great significance.
This results in a very weak coupling of two inductors when placed on two opposite sides of a closed core.
Nevertheless, the coupling is sufficient to allow the implementation of a transformer.
Future work includes to investigate the error in the inductance of model C and the influence of the magnetic fields on any underlying ICs.
Ultimately, design rules for the usage of bond wire inductors can be derived from the field results. 

\section*{Acknowledgment}
The authors thank Victoria Heinz, Isabel Kargar, David Riehl, and Timo Oster for their passionate work setting up the prototypes and simulation models.
This work is supported by the Excellence Initiative of the German Federal and State Governments and the Graduate School CE at Technische Universität Darmstadt.

\end{document}